\documentclass[12pt]{article}

\usepackage{graphicx}
\usepackage{setspace}
\usepackage{bm}

\usepackage{graphicx}

\usepackage{epic,eepic}
\usepackage{graphics}
\usepackage{epsfig}
\usepackage{pifont}
\usepackage{subfigure}
\usepackage{amsmath}

\newcommand{\ee}{\end{equation}}
\newcommand{\be}{\begin{equation}}

\begin{document}
\title{Boundary induced non linearities at small Reynolds Numbers}
\author{M. Sbragaglia {$^{1}$} and K. Sugiyama {$^{1}$} \\ 
{\small $^{1}$ Department of Applied Physics, University of Twente}\\ 
{\small P.O. Box 217, 7500 AE, Enschede, The Netherlands.}}


\maketitle
\begin{abstract}{
We investigate the influence of boundary slip velocity in Newtonian fluids at finite  Reynolds numbers. Numerical simulations with Lattice Boltzmann method (LBM) and Finite Differences method (FDM) are performed to quantify the effect of heterogeneous boundary conditions on the integral and local properties of the flow. Non linear effects are induced by the non homogeneity of the boundary condition and change the symmetry properties of the flow inducing an overall mean flow reduction. To explain the observed drag modification, reciprocal relations for stationary ensembles are used, predicting a reduction of the mean flow rate from the creeping flow to be proportional to the fourth power of the friction Reynolds number. Both numerical schemes are then validated within the theoretical predictions and reveal a pronounced numerical efficiency of the LBM with respect to FDM.}
\end{abstract}
\maketitle

\section{Introduction}

The growing interest in fluid properties and mass transfer at the micro and nanoscale \cite{hotai,tabebook}  has recently produced new research themes and questions. Above all, the clear understanding of  interfacial phenomena and wetting properties \cite{schnell,vino}, roughness effects \cite{richardson,barrat,wata} and surface nanobubbles \cite{Tyrrel} (see also \cite{laugarev} for an exhaustive review) is constantly providing new perspectives on how to interpret boundary conditions for fluid flows confined to the micro-scale and even below. In fact, the failure of the classical no-slip boundary condition in hydrodynamics \cite{Goldstein}, is  now predicted by a series of experiments \cite{pit,cheng,zhu,bonna} and numerical studies \cite{troian,barrat,koplik}.
An appealing explanation for this observed slippage is the formation of gas pockets (bubbles) between the liquid and the solid \cite{Degennes}, in such a way that they can provide  a zero shear stress condition for the flow and modify considerably its friction properties. One would then be interested to understand which are the microscopic parameters at the onset of this slip motion and that, more generally, determine the boundary stability of these bubbles. Such an ab-initio study should correctly be addressed in Molecular Dynamics simulations (MD) \cite{troian,barrat} where the problem is attacked from the atomistic point of view, integrating Newton's equation for a  set of molecules interacting through Lennard-Jones potentials. The main drawback of this approach is (by definition) its inability to describe spatial fluctuations larger than the intermolecular potential and temporal scales larger than a few milliseconds. This obviously prevents a clear understanding  of the slip effects on the flow properties on larger scales, an issue that must be addressed through continuum \cite{Stonelauga,Philip1,Philip2,Cottin} description based on the Navier-Stokes equations with slip boundary conditions or, alternatively,  with recently used mesoscopic methods based on the Boltzmann equation (the Lattice Boltzmann method (LBM)) with kinetic boundary conditions \cite{JFMsbraga,succi,AK}. More precisely, in order to study the correct continuum momentum balance one should work on the continuity and Navier-Stokes equations:
$${\bm \nabla} \cdot {\bm u}=0 $$
$$\partial_t {\bm u}+{\bm \nabla}\cdot({\bm u}{\bm u}) = -\frac{\nabla p}{\rho} + \nu \Delta {\bm u}$$
where ${\bm u}$ is the velocity vector, $p$ the internal pressure, 
$\rho$ the fluid density and $\nu$ the kinematic viscosity. These equations should then be provided with some ad hoc boundary conditions with the role of renormalizing the microscopic details of the fluid interactions  at the wall \cite{koplik,laugarev}. When such a continuum description is involved, the first non trivial control parameter on the flow is the {\em Reynolds Number}, 
$$Re=\frac{L U}{\nu}$$ 
expressing the ratio between the importance of advective terms with respect to viscous ones, being $L,U,\nu$ a typical macroscopic length, macroscopic velocity and fluid viscosity respectively. Small scale systems imply  small $Re$ ($Re \ll 1$), the consequence being that non linear terms are considered negligible and consequently ruled out. This means that when the boundary conditions are declared, we have to face a linear problem that in some cases is feasible from the mathematical point of view. In particular, the effective slip properties of pressure driven Stokes flows over heterogeneous surfaces made up of mixed  no slip and free shear walls (hereafter NSW and FSW) have been studied 
$$\nu \Delta {\bm u}= \frac{1}{\rho}{\bm \nabla}p +\frac{1}{\rho} \left( \frac{d P}{d x}\right) {\bm e}_x$$
$${\bm u}_{||} =0\hspace{.1in} \mbox{\scriptsize(NSW)}$$
$$\partial_n {\bm u}_{||} =0 \hspace{.1in} \mbox{\scriptsize (FSW)}$$
being $\frac{d P}{d x}$ an external pressure pumping (required to obtain a nonzero mass flow rate) in the direction ${\bm e}_x$, ${\bm u}_{||}$ the tangential component of the fluid velocity at the wall and $n$ the wall normal directed outward in the bulk fluid. The physical idea is to translate the presence of gas pockets in a suitable boundary condition for the hydrodynamic fields and then study the drag modifications in the fluid as a function of the degrees of freedom describing the surface heterogeneities. This problem has been addressed for the first time by Philip nearly $30$ years ago \cite{Philip1,Philip2} for the case of longitudinal strips of free shear (with respect to the flow direction). Using conformal mapping Philip related the mass flow rate gain due to the slip motion to the geometry of the boundary condition. Only more recently Stone and Lauga \cite{Stonelauga} have proposed a similar approach for pressure driven Stokes flows with transversal strips of free shear. 
In these models, non linear terms are supposed very small (formally $Re \rightarrow 0$) but, due to the surface heterogeneities, it is not clear if and how they influence the fluid by changing $Re$. In fact, due to the non homogeneity in the boundary condition, non subtle effects may be induced by the boundary and make the non linear terms play a role, an effect that would have no counterpart in laminar homogeneous flows. To clarify this point, in this paper we will study the overall sensitivity of the fluid with respect to these effects. We will carry out a complete study of pressure driven laminar flows with mixed boundary conditions of free shear and no slip for {\em finite Reynolds numbers} :
$${\bm \nabla}\cdot({\bm u}{\bm u})= \nu \Delta {\bm u}-\frac{1}{\rho}{\bm \nabla}p-\frac{1}{\rho} \left( \frac{d P}{d x}\right) {\bm e}_x$$ 
$${\bm u}_{||} =0\hspace{.1in} \mbox{\scriptsize(NSW)}$$
$$\partial_n {\bm u}_{||} =0 \hspace{.1in} \mbox{\scriptsize (FRW)}.$$
 Numerical Simulations with Finite Difference methods (FDM) and Lattice Boltzmann methods (LBM) are used firstly to check the analytical estimates proposed by Stone and Lauga (for $Re=0$) in the case of transversal strips  and then to characterize the flow properties at $Re>0$. Interesting flow behaviors are observed for finite $Re$ and a non trivial interplay between the boundary condition and non linear terms is considered both theoretically and numerically. In particular, the mass flow rate, $\langle \tilde{u} \rangle$, for finite Reynolds will differ from its creeping flow ($Re=0$) counterpart, $\langle u \rangle$, in an exact way:
$$\langle u\rangle-\langle \tilde{u}\rangle
=
\langle ({\bm u}{\bm u}):({\bm \nabla}\tilde{\bm u})\rangle.$$
Employing finite Reynolds perturbation theory, for small Reynolds numbers, we will show that the observed drag modification, even if triggered by the boundary, has a scaling law behavior with respect to the bulk Reynolds number
$$\langle u \rangle/\langle \tilde{u} \rangle -1 \sim Re^{2}.$$
In order to validate this scaling law behavior and its range of applicability, all the observed effects are quantitatively studied with numerical simulations. The result is that up to $Re \sim 1$ the predicted scaling law behavior is satisfied with high accuracy in both numerical schemes and the friction properties of the system are influenced by the presence of non linear terms in such a way that the flow differs from its creeping flow counterpart both qualitatively and quantitatively. In fact soon after non linear effects are induced by the boundary conditions, the flow is still laminar but its local symmetry properties change as an overall drag enhancement is induced.\\
The paper is organized as follows: in section \ref{sec2} the physical problem is mathematically formulated and we will give the correct background to apply finite Reynolds perturbation theory that is the subject of section \ref{sec3}. In section \ref{sec4} we give a brief review of the numerical procedures used. All numerical results are discussed in section \ref{sec5} and conclusions will follow in section \ref{sec6}.

\section{Formulation of the problem}\label{sec2}

In this section we formulate the problem under consideration from the mathematical point of view. We will refer to a wall-bounded flow where $x$, $y$ and $z$ denote respectively the streamwise, spanwise and wall-normal coordinates. The governing equations are expressed in the dimensional form:
\be
{\bm \nabla}^{*}\cdot {\bm u}^{*}=0,
\ee
\be
{\bm \nabla}^{*}\cdot({\bm u}^{*} {\bm u}^{*})
=
{\bm \nabla}^{*}\cdot{\mbox{\boldmath$\sigma$}^{*}}
+\frac{1}{\rho^{*}}\left(-\frac{{\rm d}P^{*}}{{\rm d}x^{*}}\right){\bm e}_x,
\ee
where ${\bm u^{*}}(=(u^{*}, v^{*}, w^{*}))$ is the velocity vector,  $\mbox{\boldmath$\sigma^{*}$}$ the stress tensor, $\rho^{*}$ the density of fluid, $-{{\rm d}P^{*}}/{{\rm d}x^{*}}$ the driving pressure gradient,  and ${\bm e}_x$ the unit vector in the $x-$direction. In the above equations, the superscript $*$ has been introduced in order to indicate dimensional quantities and to distinguish them from dimensionless quantities (without superscript) to be introduced later.\\
For the incompressible Newtonian fluid,  the stress tensor is written as
\be
\mbox{\boldmath$\sigma^{*}$}=
-\frac{p^{*}}{\rho^{*}}{\bm I}+2\nu^{*}{\bm S}^{*},
\ee
where $p^{*}$ is the deviation of the pressure from the driving component, 
${\bm I}$ the unit dyadic, 
$\nu^{*}$ the kinematic viscosity and ${\bm S}^{*}$ the strain rate
given by
\be
{\bm S}^{*}=\frac{1}{2}\left\{ {\bm \nabla}^{*}{\bm u}^{*}+({\bm \nabla}^{*}{\bm u}^{*})^{\rm T}\right\}.
\ee
Boundary conditions  on the no-slip wall (NSW) and  the free slip wall (FSW) are written as
\be
{\bm u}^{*}=0\ \ \ {\rm on}\ \ \ {\rm NSW},
\ee
\be
({\bm n}\cdot\mbox{\boldmath$\sigma$}^{*})\times {\bm n}=0
\ \ \ {\rm on}\ \ \ {\rm FSW},
\ee
where ${\bm n}$ is the wall-normal unit vector. Two walls are then considered at $z=0$ and $z=L_z$ while
in the streamwise and  spanwise directions, 
periodic boundary conditions are imposed 
with the period length of $L_x$ and $L_y$. Note that the relation of 
${\bm n}\cdot\mbox{\boldmath$\sigma$}^{*}\cdot{\bm u}^{*}=0$
is satisfied on both the no slip and free shear walls. 
In the present study, we use symmetric boundary conditions in the upper and lower walls and we will refer to the slip percentage, $\xi$, as the ratio between the free shear area with respect to the total one.\\
To proceed further, we should introduce dimensionless equations to which we will apply finite Reynolds perturbation theory.  To do so,  let us  now consider the friction velocity $u_\tau$ and  the friction Reynolds number $Re_\tau$ defined as
\be
u_\tau=\sqrt{\tau_w},
\ee
\be\label{friction}
Re_\tau=\frac{u_\tau(L_z/2)}{\nu},
\ee
where $\tau_w$ is the skin friction averaged over the wall surface at $z=0$ (or $z=L_z$). Since  symmetric boundary conditions are imposed on $z=0$ and $z=L_z$, we obtain the relation between  the skin friction and the driving pressure gradient:
\be
\tau_w=\nu \left\langle
\frac{\partial u}{\partial z}
\right\rangle_{z=0}=-\nu
\left\langle
\frac{\partial u}{\partial z}
\right\rangle_{z=L_z}
=
\frac{L_z}{2\rho}
\left(-\frac{{\rm d}P}{{\rm d}x}\right)
\ee
that immediately imply:
\be\label{ReRe}
Re \propto Re^2_{\tau}.
\ee
Now, we choose $u_\tau$ and $L_z/2$ as the velocity and length scales for normalization: 
\be
{\bm u}=(u,v,w)=(u^{*}/u_{\tau},v^{*}/u_{\tau},w^{*}/u_{\tau})
\ee
\be
{\bm x}=(x,y,z)=(2 x^{*}/L_z,2 y^{*}/L_z,2 z^{*}/L_z).
\ee
This means that the two walls are located at $z=0$ and $z=2$ and  that the  velocity components and the pressure fields satisfy the following dimensionless equations:
\be\label{dimensionless1}
{\bm \nabla}\cdot{\bm u}=0,
\ee
\be\label{dimensionless2}
{\bm \nabla}\cdot({\bm u}{\bm u})={\bm \nabla}{\bm \sigma}+{\bm e}_{x}
\ee
with the boundary conditions
\be
{\bm u}=0\ \ \ {\rm on}\ \ \ {\rm NSW},
\ee
\be
({\bm n}\cdot\mbox{\boldmath$\sigma$})\times {\bm n}=0
\ \ \ {\rm on}\ \ \ {\rm FSW},
\ee
where the pressure has been normalized with the term $(\rho^{*} u_\tau^{2})$
and the dimensionless stress tensor is given by 
$$
\mbox{\boldmath$\sigma$}
=-p{\bm I}+\frac{2}{Re_\tau}{\bm S}.
$$
We also introduce the velocity vector 
$\tilde{\bm u}$ and the pressure $\tilde{p}$ satisfying 
the following the equations for the creeping flow with no advection:
\be
{\bm \nabla}\cdot\tilde{\bm u}=0,
\ee
\be
0={\bm \nabla}\cdot\tilde{\mbox{\boldmath$\sigma$}}+{\bm e}_x,
\ee
with the boundary conditions
\be
\tilde{\bm u}=0\ \ \ {\rm on}\ \ \ {\rm NSW},
\ee
\be
({\bm n}\cdot\tilde{\mbox{\boldmath$\sigma$}})\times {\bm n}=0
\ \ \ {\rm on}\ \ \ {\rm FSW}.
\ee
In this case, $\tilde{\bm u}$ and $\tilde{p}$ represent the zeroth order approximation for the velocity field and stress tensor and they will be used to quantify the effects of a finite Reynolds number. Our starting point is the application of reciprocal relations \cite{lorentz} to the fields ${\bm u}$ and $\tilde{\bm u}$ and their stresses. In fact it can be shown that
\be\label{rec}
({\bm \nabla}{\bm u}):\tilde{\mbox{\boldmath$\sigma$}} = ({\bm \nabla}\tilde{\bm u}):\mbox{\boldmath$\sigma$}
\ee
is an exact relation between the above fields satisfied in every point of the space. Taking the volume integral of the lhs of (\ref{rec}) over the whole fluid region,  we obtain 
\begin{equation}\label{integral1}
\begin{split}
\int_{\mbox{\footnotesize fluid}}\!\!\!{\rm d}^3{\bm x}\
({\bm \nabla}{\bm u}):\tilde{\mbox{\boldmath$\sigma$}}
=&
-\int_{\mbox{\footnotesize fluid}}\!\!\!{\rm d}^3{\bm x}\ 
{\bm u}\cdot({{\bm \nabla}\cdot\tilde{\mbox{\boldmath$\sigma$}}})
+
\int_{\mbox{\footnotesize fluid}}\!\!\!{\rm d}^3{\bm x}\ 
{\bm \nabla}\cdot ({\bm u}\cdot \tilde{\mbox{\boldmath$\sigma$}})
\\
=&
\int_{\mbox{\footnotesize fluid}}\!\!\!{\rm d}^3{\bm x}\ 
{\bm u}\cdot{\bm e}_x
+
\oint_{\mbox{\footnotesize wall}}\!\!\!{\rm d}^2{\bm x}\ 
\underbrace{
{\bm n}\cdot \tilde{\mbox{\boldmath$\sigma$}}\cdot{\bm u}
}_{=0}\\
&=L_xL_zL_z\langle u\rangle,
\end{split}
\end{equation}
where $\langle \cdot\rangle$ represents the volume average operator. 
In the similar way, for the rhs of (\ref{rec}), we obtain 
\begin{equation}\label{integral2}
\begin{split}
\int_{\mbox{\footnotesize fluid}}\!\!\!{\rm d}^3{\bm x}\ 
({\bm \nabla}\tilde{\bm u}):\mbox{\boldmath$\sigma$}
=&
\int_{\mbox{\footnotesize fluid}}\!\!\!{\rm d}^3{\bm x}\ 
\tilde{\bm u}\cdot{\bm e}_x
-
\int_{\mbox{\footnotesize fluid}}\!\!\!{\rm d}^3{\bm x}\ 
\tilde{\bm u}\cdot\left\{({\bm u}\cdot{\bm \nabla}){\bm u}\right\}
+
\oint_{\mbox{\footnotesize wall}}\!\!\!{\rm d}^2{\bm x}\ 
\underbrace{
{\bm n}\cdot {\mbox{\boldmath$\sigma$}}\cdot\tilde{\bm u}
}_{=0}\\
=&L_xL_zL_z\langle \tilde{u}\rangle
-
\int_{\mbox{\footnotesize fluid}}\!\!\!{\rm d}^3{\bm x}\ 
\tilde{\bm u}\cdot\left\{({\bm u}\cdot{\bm \nabla}){\bm u}\right\}.
\end{split}
\end{equation}
where the second term in the rhs is 
\begin{equation}\label{integral3}
\begin{split}
-\int_{\mbox{\footnotesize fluid}}\!\!\!{\rm d}^3{\bm x}\ 
\tilde{\bm u}\cdot\left\{({\bm u}\cdot{\bm \nabla}){\bm u}\right\}
=&
\int_{\mbox{\footnotesize fluid}}\!\!\!{\rm d}^3{\bm x}\ 
{\bm \nabla}\cdot\left\{{\bm u}({\bm u}\cdot\tilde{\bm u})\right\}
+
\int_{\mbox{\footnotesize fluid}}\!\!\!{\rm d}^3{\bm x}\ 
({\bm u}{\bm u}):({\bm \nabla}\tilde{\bm u})
\\
=&
\oint_{\mbox{\footnotesize wall}}\!\!\!{\rm d}^2{\bm x}\ 
\underbrace{
{\bm n}\cdot\left\{{\bm u}({\bm u}\cdot\tilde{\bm u})\right\}
}_{=0}
+
L_xL_yL_z\langle ({\bm u}{\bm u}):({\bm \nabla}\tilde{\bm u})\rangle.
\end{split}
\end{equation}
Using (\ref{integral1}),(\ref{integral2}) and (\ref{integral3})
we then obtain the contribution of the advection to 
the change of the flow rate
\be\label{master}
\langle u\rangle-\langle \tilde{u}\rangle
=
\langle ({\bm u}{\bm u}):({\bm \nabla}\tilde{\bm u})\rangle.
\ee
Note that this expression is consistent with 
the identity of the Reynolds shear stress contribution 
to the friction coefficient 
in a turbulent flow bounded with no-slip walls
derived in \cite{fukagata} where the authors derived the following relation 
\begin{equation}
C_f=\frac{12}{Re}+\frac{6\langle(1-z)(-\overline{u'w'})\rangle}
{\langle u \rangle^2},
\end{equation}
where $C_f$ 
is the friction coefficient, 
$z$ the wall normal coordinate normalized 
by the channel half width, $L_z/2$,
and $-\overline{u'w'}$ the Reynolds shear stress. 
This identity indicates that 
the effect of the Reynolds shear stress on the skin friction 
is weighted by the factor of $1-z$. 
Substituting 
${\bm \nabla}\tilde{\bm u}=
{\bm e}_x{\bm e}_z{Re_\tau}(1-z)$
with the no-slip walls, 
which contains the same weighted factor of $1-z$, 
into (\ref{master}) and considering 
$\langle\tilde{\bm u}\rangle=Re_\tau/3$, 
one can obtain the same relation.

\section{Finite $Re_\tau$ Perturbation theory}\label{sec3}

Using the perturbation method with respect to
the friction Reynolds number $Re_\tau$ 
under the assumption of $0<Re_\tau\ll 1$, 
the effect of the advection on 
the change of the flow rate
$\langle u\rangle-\langle \tilde{u}\rangle$
is now examined. More precisely, we do not directly solve the perturbation equation but we want to consider the following relation  
$$\langle u\rangle-\langle \tilde{u}\rangle=\Gamma Re_\tau^n$$ 
(here $\Gamma$ is a coefficient) for  small Reynolds numbers and concentrate on the exponent $n$. 
In our present approach, ${\bm u}$ and $p$ in (\ref{dimensionless1}) and (\ref{dimensionless2}) are regularly expended  with respect to $Re_\tau$ in the following form
\be
{\bm u}=
Re_\tau^1{\bm u}^{(0)}
+
Re_\tau^3{\bm u}^{(1)}
+
Re_\tau^5{\bm u}^{(2)}+...,
\ee
\be
p=
Re_\tau^0 p^{(0)}
+
Re_\tau^2 p^{(1)}
+
Re_\tau^4 p^{(2)}+...,
\ee
where the properties with the overscript $(0)$, 
indicate the creeping flow solution,  
\be
\tilde{\bm u}=Re_\tau{\bm u}^{(0)},
\ \ \ 
\tilde{p}=p^{(0)}.
\ee
Note that the even order  (i.e., $O(Re_\tau^{2n})$)
components of the velocity vector are identically zero
because there is not any advection or driving forcing 
to balance with the divergence of the stress tensor 
containing the $O(Re_\tau^{2n})$ velocity.  A detailed calculation is shown in Appendix with the final result that the change in the flow rate is such that
$$\langle u\rangle-\langle \tilde{u}\rangle =\Gamma Re_\tau^5,$$ 
with the prefactor 
$$\Gamma =
\left
\langle 
{\bm u}^{(1)}\cdot
\left\{
({\bm u}^{(0)}\cdot {\bm \nabla}){\bm u}^{(0)}
\right\}
\right\rangle$$ 
that is dependent upon the perturbed velocity component
but independent of the Reynolds number. Moreover,
since $\langle \tilde{u}\rangle=Re_\tau\langle u^{(0)}\rangle$
is proportional to $Re_\tau$, 
the relative change of the flow rate is such that
\be\label{powlaw}
\langle u\rangle/\langle \tilde{u}\rangle-1 \sim Re^{4}_{\tau}.
\ee
Alternatively if one is interested in a scaling relation with respect to the bulk Reynolds number, it follows immediately from (\ref{ReRe}) that 
\be
\langle u\rangle/\langle \tilde{u}\rangle-1\sim Re^2.
\ee
Note that this effect is physically induced by the boundary condition ($\Gamma >0$) but soon after $Re_{\tau}>0$ it develops with very well defined scaling laws with respect to the Reynolds number. Note also that for any finite Reynolds number, we should expect the velocity to become uniformly lower  than that in the creeping flow. In fact, according to Lamb \cite{Lamb}, 
the point in which the velocity field takes the minimum of 
the total dissipation rate is the solution of the creeping flow 
with the prescribed boundary condition. 
If the flow rate is fixed, 
the total dissipation rate is proportional to 
the driving pressure gradient, 
which equals the skin friction divided 
by the channel half width. This means that the deviation of the velocity from the creeping flow due to the advective motion of fluid
enhances the total skin friction. 
It can be thus interpreted that a larger driving force 
than that in the creeping flow
is required at the finite Reynolds number
in order to maintain a given flow rate. Equivalently, 
with fixed driving pressure gradient and viscosity, 
the flow rate should reduce due to the advective motion.
Therefore, 
$\langle u\rangle-\langle \tilde{u}\rangle$
is negative at the non-zero Reynolds number, that is equivalent to 
$\left
\langle 
{\bm u}^{(1)}\cdot
\left\{
({\bm u}^{(0)}\cdot{\bm \nabla}){\bm u}^{(0)}
\right\}
\right\rangle<0$
for $Re_\tau>0$.

\section{Numerical Procedures}\label{sec4}

In order to validate the previous theoretical analysis we will use  standard Finite Difference methods (FDM) and Lattice Boltzmann methods (LBM). \\
In the Finite Difference Method, the equations are discretized in an Eulerian framework on the staggered grid \cite{har}.
The second-order scheme, 
i.e., the Adams-Bashforth method for the advection term
and the Crank-Nicolson one for the viscous term,
is used to integrate the equations in time \cite{can}.  
The pressure is treated implicitly. 
The space derivatives are approximated 
by the fourth-order central difference scheme. 
In particular for the advection term, 
we employ the scheme by 
Kajishima {\it et al.} in \cite{kaji}.\\
In solving the Poisson equation, 
we use the Fast Fourier Transform (FFT)
for the high speed and accuracy. 
The three-dimensionally discretized equations 
are then reduced into the one-dimensional problem 
by taking the FFT 
in the homogeneous (streamwise and spanwise) directions. 
The reduced-order equation written in the heptadiagonal matrix form 
is directly solved. Since we use the staggered grid system, 
the velocity parallel to the wall ${\bm u}_{||}$ 
is not located on the wall. In order to approximate the boundary conditions on the wall, 
the velocity at the virtual point outside the flow region is adjusted 
using the third-order Lagrange extrapolation.\\ 
The other numerical technique used here is a  mesoscopic approach based on the Boltzmann equations and known as Lattice Boltzmann Method (LBM) \cite{saurobook,gladrow}. This method is a kinetic approach to fluid flows that starts from the Boltzmann equation in discrete form
\be\label{LBM}
f_{i}({\bm x}+\Delta t{\bm c}_i,t+\Delta t)-f_{i}({\bm x},t)=-\frac{\Delta t}{\tau} (f({\bm x},t)-f^{(eq)}({\bm x},t))+F_{i}.
\ee
The left hand side represents a streaming term of a probability density function ($f_i({\bm x},t)$) to find in a given position and time (${\bm x},t$) a kinetic particle whose velocity is ${\bm c}_i$. Here the set of velocities is properly discretized (${\bm c}_i, i=0,...,N$) dependently on the symmetry properties of the lattice. The right hand side represents a simplified version of the standard collision term in the real Boltzmann equation: it expresses a relaxation (with characteristic time $\tau$) towards a local equilibrium ($f^{(eq)}$), the equivalent of the local Maxwellian equilibrium in kinetic theory \cite{cerci}. Finally the term $F_i$ is an external volume forcing, used to produce a constant pressure drop ($dP/d x$) in the streamwise direction. Starting from the kinetic equations (\ref{LBM}) and coarse-graining (in the velocity space) the kinetic distributions, one obtains respectively a macroscopic density and velocity field as follows:
\be\label{den}
\rho({\bm x},t)=\sum_{i}f_i({\bm x},t)
\ee
\be\label{vel}
{\bm u}({\bm x},t)=\frac{1}{\rho({\bm x},t)}\sum_{i}{\bm c}_if_i({\bm x},t).
\ee
Typically, to connect the above macroscopic fields to the continuum description of Navier-Stokes equations one needs  a separation of scale parameter, usually the Knudsen Number
\be
Kn=\lambda/L
\ee
being $\lambda \sim |c_s \tau|$ a length scale of the order of the mean free path of the kinetic particles (here $c_s$ denotes the sound speed velocity), and $L$ a macroscopic length. In the limit $Kn \ll 1$ it can be shown \cite{gladrow,chen} that the macroscopic fields (\ref{den}) and (\ref{vel}) satisfy the continuum equations of fluid mechanics:
\be
\partial_t \rho+ {\bm \nabla} \cdot (\rho {\bm u})=0
\ee
\be
\rho[\partial_t {\bm u}+({\bm u} \cdot {\bm \nabla}){\bm u}]=-{\bm \nabla}p+\rho \nu \Delta {\bm u}-\frac{d P}{d x} {\bm e}_x 
\ee
being $p$ the internal pressure and  $\nu$ the fluid viscosity, directly related to the relaxation time of the Boltzmann equation, $\nu=c^2_s(\tau-\Delta t/2)$ (see also \cite{gladrow} for an exhaustive review of the method).\\
The subject of boundary condition for the Lattice Boltzmann Method has been extensively debated in the past years. An adequate set of kinetic boundary conditions has been proposed in \cite{AK} through a lattice transcription of the accommodation coefficients used in continuum kinetic theory of rarefied gases \cite{cerci2} and later similar kinetic boundary condition approaches have been proposed on more empirical grounds \cite{succi,sbraga,emerson}. In particular in \cite{JFMsbraga} it has been shown that the introduction of a  {\em slip function}  at the kinetic level is enough to produce, for small Knudsen numbers, local macroscopic boundary conditions of the form:
\be
{\bm u}_{||}({\bm x}) \sim Kn \frac{s({\bm x})}{1-s({\bm x})} \partial_{n} {\bm u}_{||} ({\bm x}).
\ee
The slip function $s({\bm x})$ represents a local degree of slippage in ${\bm x}$, tunable from pure no-slip properties ($s=0$) up to free shear condition ($s=1$). This local and tunable slip properties has already been used in \cite{JFMsbraga} to match the macroscopic calculation proposed by Stone and Lauga in \cite{Stonelauga} and Philip in \cite{Philip1,Philip2} for boundary condition made up of strips of free shear and no-slip.

\section{Numerical Results}\label{sec5}

\begin{figure}[t!]
\begin{center}
\includegraphics[scale=0.6]{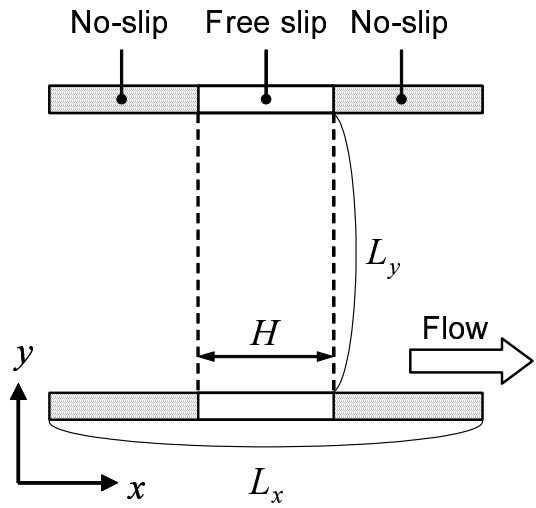}
\end{center}
\caption{The computational scheme for the numerical simulations. We use a $2d$ channel with the same boundary conditions on the top and lower walls. Free shear conditions are imposed on a width $H$ on both top and lower walls giving a slip percentage $\xi=H/L_{x}$. We also impose periodic boundary conditions in the streamwise ($x$) direction.}
\label{fig:01}
\end{figure}

\begin{figure}[t!]
\begin{center}
\includegraphics[scale=0.6,angle=-90]{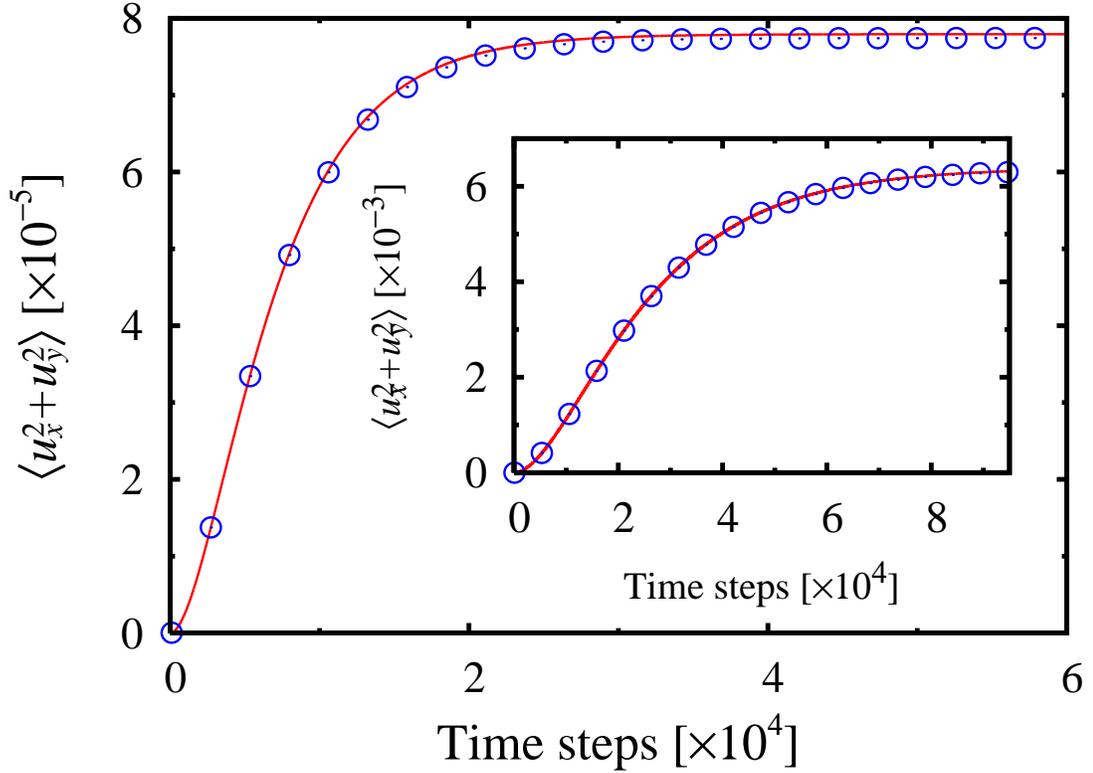}
\end{center}
\caption{Time evolution of the total kinetic energy, $\langle u^2_{x}+u^2_{y}\rangle$, as a function of time steps. In the main figure we show a case with friction Reynolds number $Re_{\tau}=2.245$ and parameters $L_{x}=64$,$L_{y}=84$,$\nu=1/6$,$-dP/dx=1.889 10^{-6}$ and $\xi=0.5$. The agreement between LBM ($\circ$) and FDM (straight lines) is very good and at the same time the LBM converges more rapidly ($\sim 5$ times faster). Inset: we report the same kind of plot for the case $Re_{\tau}=10.04$, $L_{x}=64$,$L_{y}=168$,$\nu=1/6$,$-dP/dx=4.72 10^{-6}$ and $\xi=0.5$. Again LBM and FDM agree very well and in this case LBM is up to $15$ times faster than FDM.}
\label{fig:02}
\end{figure}

\begin{figure}[t!]
\begin{center}
\includegraphics[scale=0.6,angle=-90]{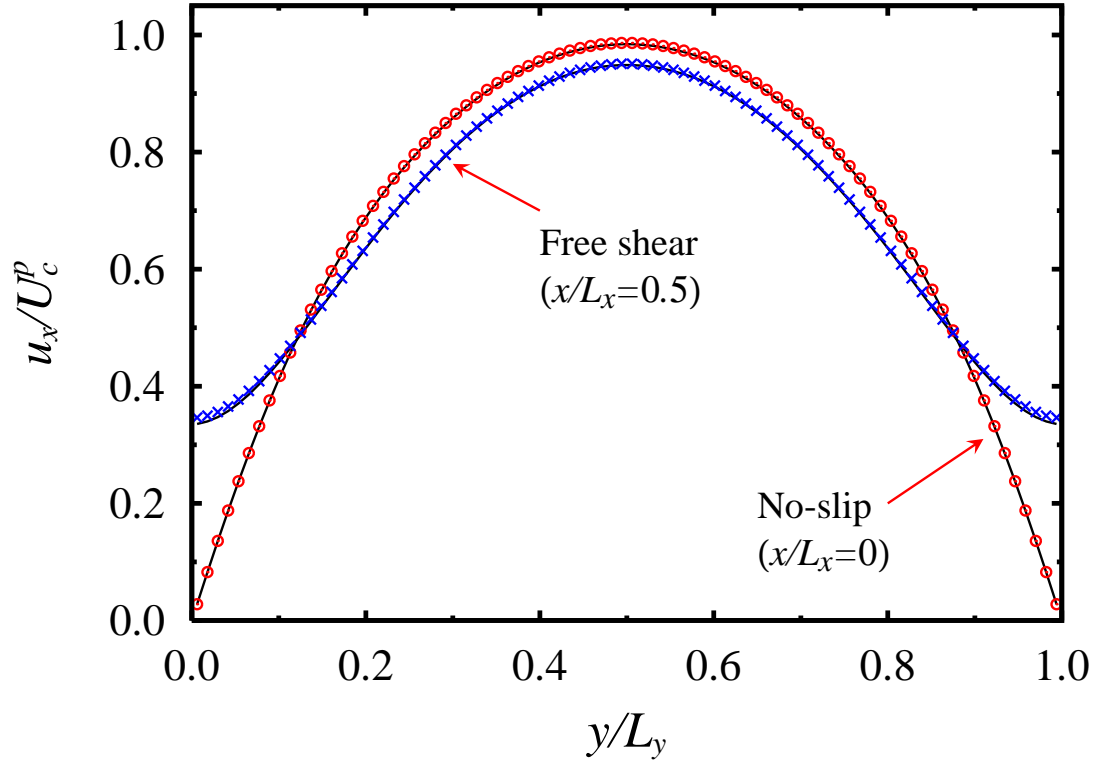}
\end{center}
\caption{Velocity profiles along a channel with mixed boundary conditions of free shear and no-slip. In the inlet region ($\circ$ for LBM and straight lines for FDM), due to the no slip condition the profile is almost a parabolic one with zero velocity at the boundaries and constant concavity, as predicted from the Poiseuille solution. In the middle of the channel ($\times$ for LBM and straight lines for FDM), the free shear condition is such to produce an enhancement of the local slip properties at the boundaries with a reduction of the profiles in the bulk region. In both cases all the profiles are normalized with respect to the center channel velocity in the Poiseuille limit.}
\label{fig:03}
\end{figure}

\begin{figure}[t!]
\begin{center}
\includegraphics[scale=0.6,angle=-90]{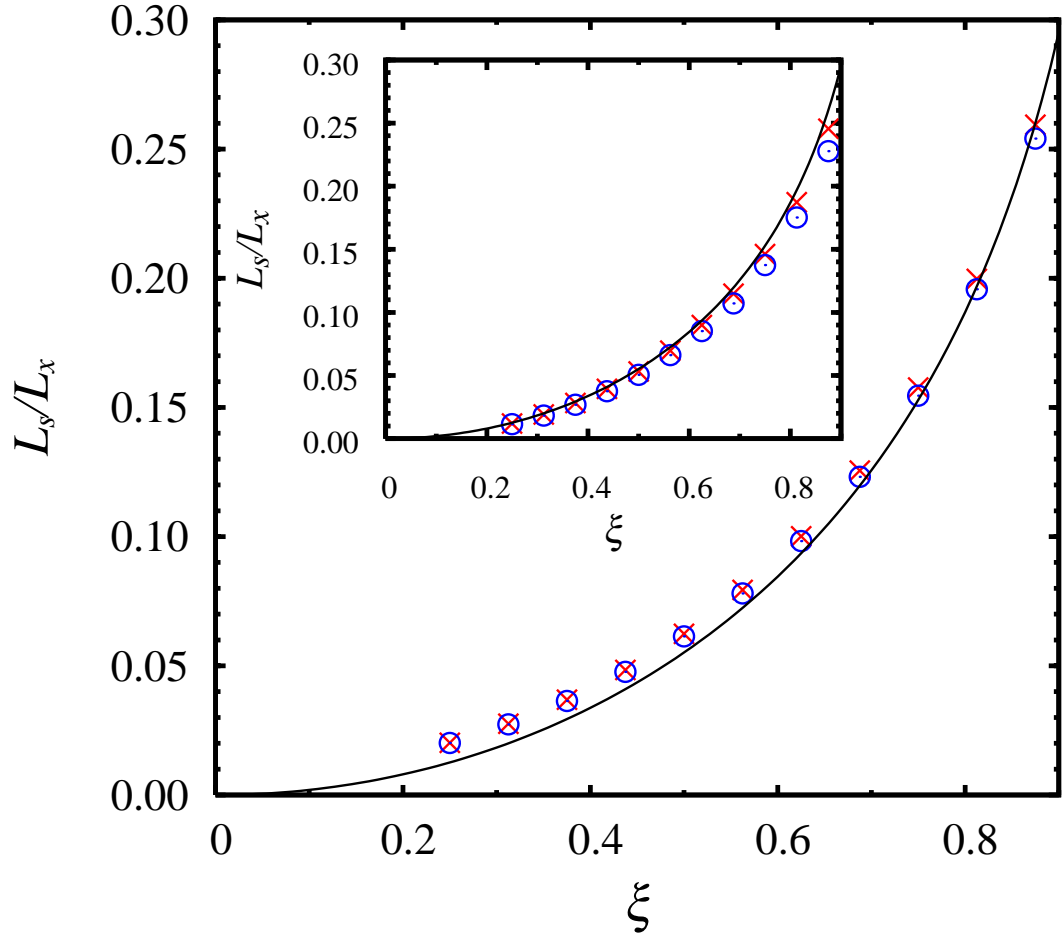}
\end{center}
\caption{Slip length normalized to the pattern dimension as a function of the slip percentage $\xi$. In the main figure we show a first order extrapolation  for LBM ($\circ$) and FDM ($\times$). Both numerical schemes are compared with the analytical estimate (straight line) given in \cite{Stonelauga}. The numerical simulations are carried out with the following set of parameters  $Re_{\tau}=2.245$,$L_{x}=64$,$L_{y}=84$,$\nu=1/6$,$-dP/dx=1.889 10^{-6}$.Inset: the same as the main figure but for the case of a fourth order approximation. In Both figures, being the Reynolds number small but finite, small discrepancies  are observed between the numerics and the analytical results.}
\label{fig:05}
\end{figure}

\begin{figure}[t!]
\begin{center}
\includegraphics[scale=0.6,angle=-90]{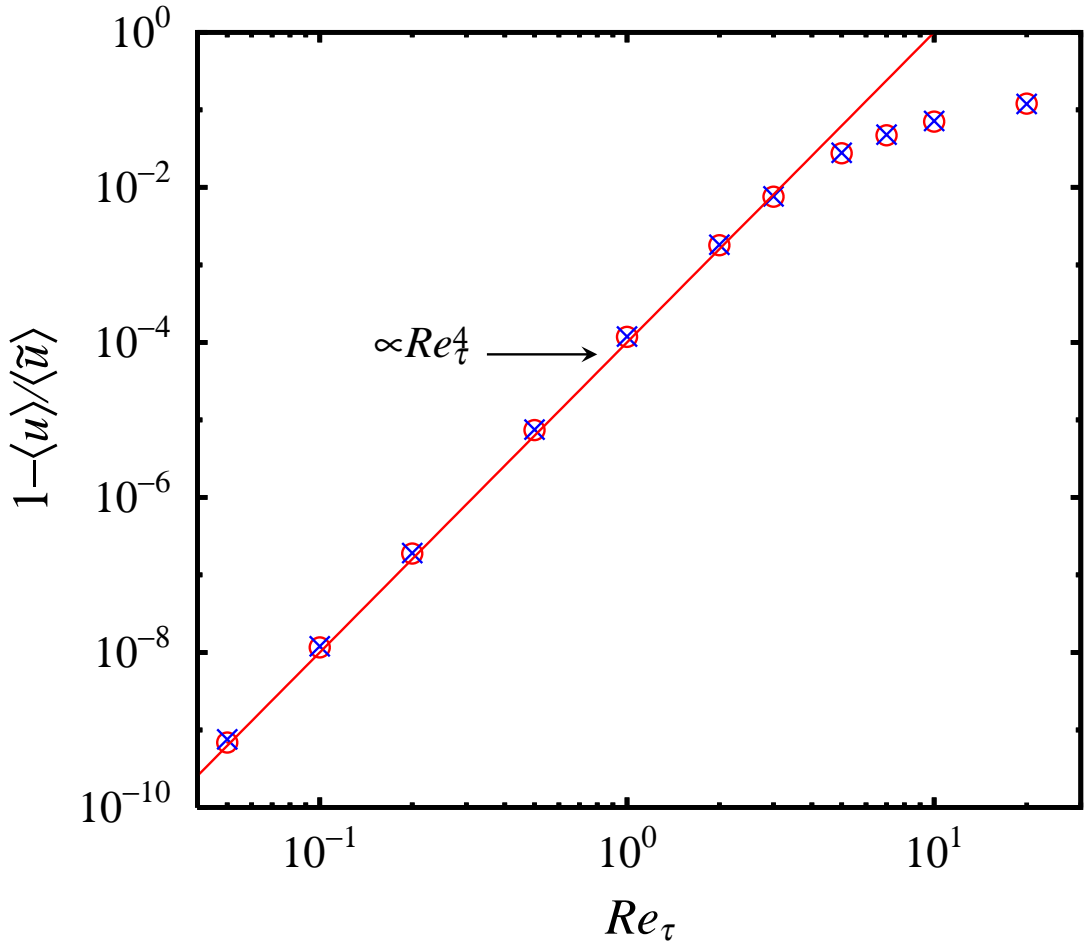}
\end{center}
\caption{Drag modification at finite Reynolds numbers. We show the relative departure from the creeping flow solution, $1-\langle u \rangle/\langle \tilde{u} \rangle$ ($\circ$), as a function of the friction Reynolds number $Re_{\tau}$. To verify the correctness of relation (\ref{master2}) we also plot the right hand side of this relation ($\times$). All the results have been obtained with FDM with numerical grid mesh $L_{x}=64$,$L_{y}=64$ and parameters $\nu=1/6$,$\xi=0.5$. The pressure gradient $dP/dx$ has then been changed to vary $Re_{\tau}$. To emphasize the scaling behavior with respect to $Re_{\tau}$, the power law $Re^4_{\tau}$ is also plotted.}
\label{fig:06}
\end{figure}

\begin{figure}[t!]
\begin{center}
\includegraphics[scale=0.6,angle=-90]{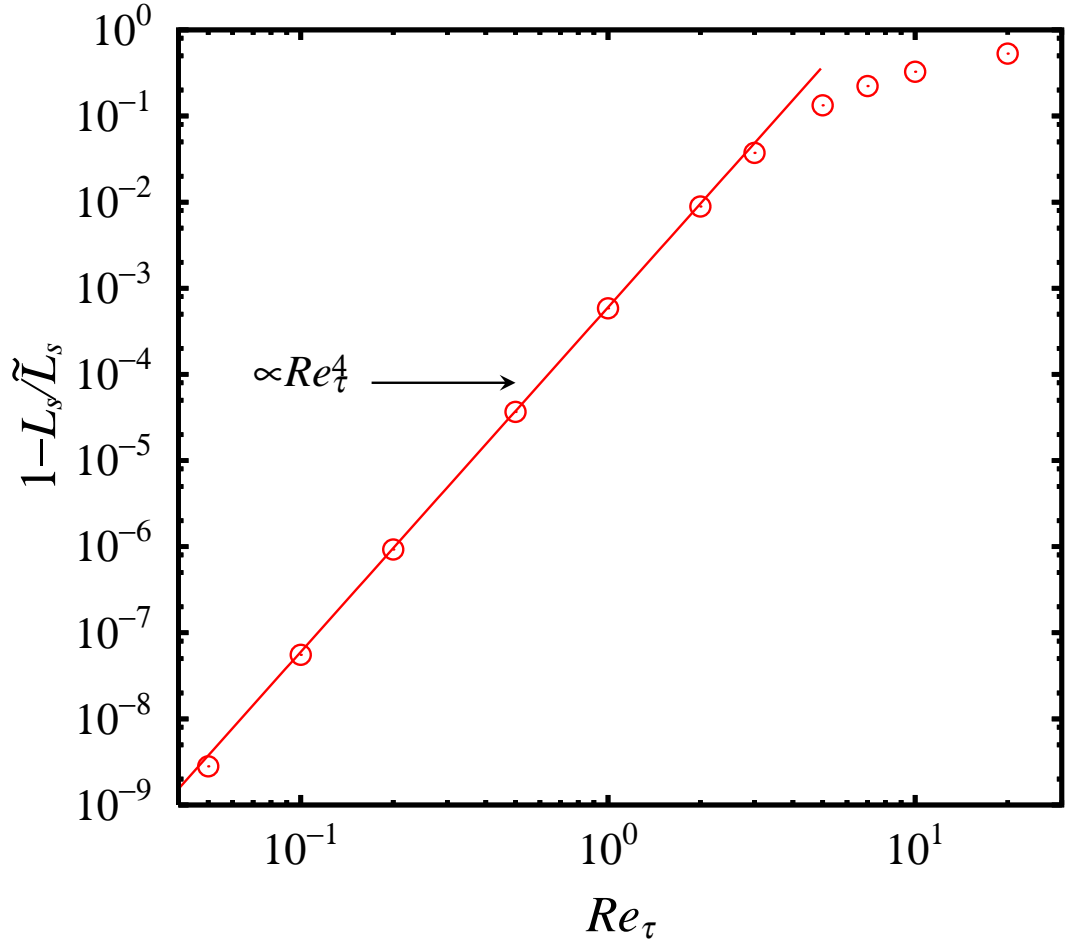}
\end{center}
\caption {Slip length at finite Reynolds numbers. We show the volume averaged slip length normalized to the creeping flow counterpart, as a function of the friction Reynolds number $Re_{\tau}$. All the results have been obtained with FDM with numerical grid mesh $L_{x}=64$,$L_{y}=64$ and parameters $\nu=1/6$,$\xi=0.5$. The pressure gradient $dP/dx$ has then been changed to vary $Re_{\tau}$. In order to highlight the scaling behavior with respect to $Re_{\tau}$, the power law behavior $Re^{4}_{\tau}$ is also represented. Note that for $Re_{\tau} \sim 10$, the overall slip length differs from its creeping flow counterpart of the order of ten percent. }
\label{fig:07}
\end{figure}

\begin{figure}[t!]
\begin{center}
\includegraphics[scale=0.6,angle=-90]{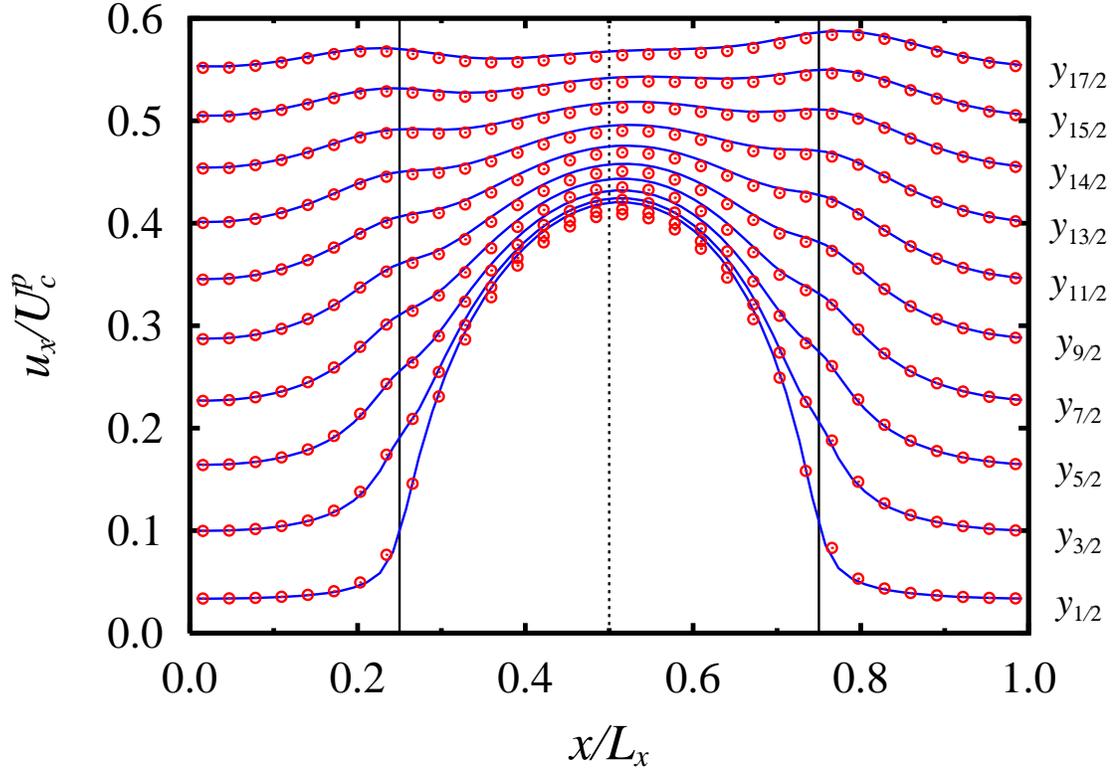}
\end{center}
\caption{Streamwise profiles for small Reynolds number. We plot the streamwise profiles as a function of the relative position  along the channel for different distances from the wall ($y_{n+\frac{1}{2}}$ with $n=0,1,2,... $ and the wall being located at $y_{0}=0$). Here we show a case with friction Reynolds number $Re_{\tau}=2.245$ and parameters $L_{x}=64$,$L_{y}=84$,$\nu=1/6$,$-dP/dx=1.889 10^{-6}$ and $\xi=0.5$.The velocity is normalized with respect to the center channel velocity of the corresponding Poiseuille profile and  both LBM ($\circ$) and FDM (straight lines) indicate an almost symmetric configuration with respect to the center of the slip area (see dotted line). This is predicted by the symmetry properties of the Stokes solution ($Re_{\tau}=0$) and it is here expected to hold with some very small corrections due to the finite Reynolds effects. }
\label{fig:08}
\end{figure}

\begin{figure}[t!]
\begin{center}
\includegraphics[scale=0.6,angle=-90]{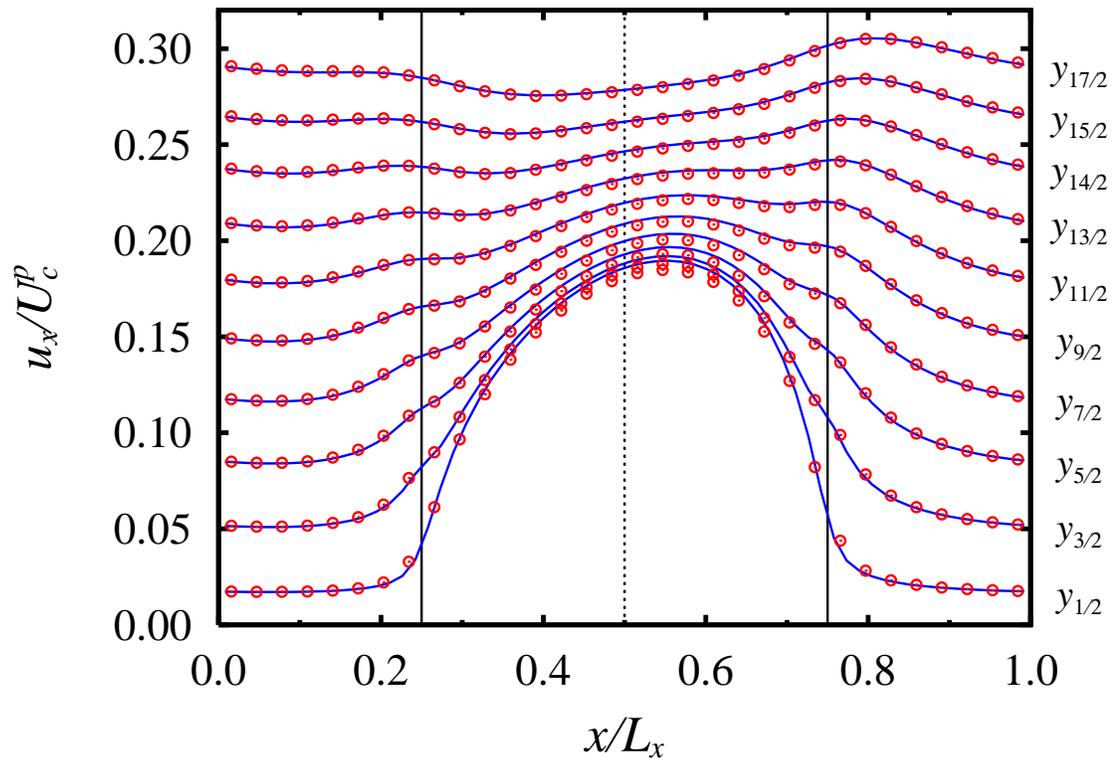}
\end{center}
\caption{Same as figure \ref{fig:08} with the following set of parameters: $Re_{\tau}=10.04$ and  $L_x=64$,$L_y=168$,$\nu=1/6$,$-dP/dx=4.72 10^{-6}$,$\xi=0.5$. Now, due to the increase of the Reynolds number the profile becomes more asymmetric with respect to the case of figure \ref{fig:08}. To highlight this effect we plot the symmetry axis of the free shear strip to be compared with the symmetry properties of the flow.}
\label{fig:09}
\end{figure}

\begin{figure}[t!]
\begin{center}
\includegraphics[scale=0.6,angle=-90]{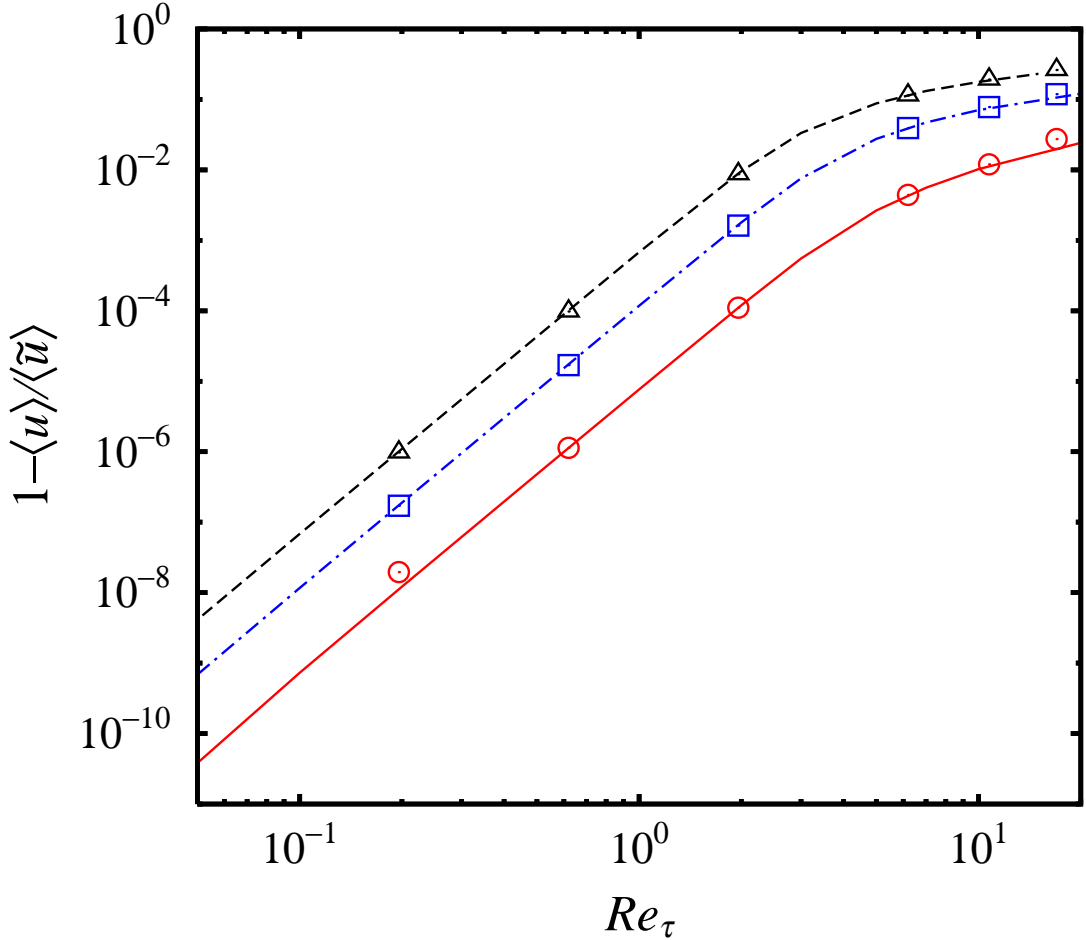}
\end{center}

\caption{Drag modification and its scaling behavior with respect to the friction Reynolds number for different values of the slip percentage $\xi$. We show the relative departure from the creeping flow solution, $1-\langle u \rangle/\langle \tilde{u} \rangle$ estimated with LBM (circles for $\xi=0.25$, squares for $\xi=0.5$ and diamonds for $\xi=0.75$) and FDM (straight lines). All the results have been obtained with FDM with numerical grid mesh $L_{x}=64$,$L_{y}=64$ and parameters $\nu=1/6$,$\xi=0.5$. The pressure gradient $dP/dx$ has then been changed to vary $Re_{\tau}$. In the small $Re_{\tau}$ regime it is observed the scaling relation proportional to $Re^4_{\tau}$ predicted by our analytical approach.}
\label{fig:11}
\end{figure}

\begin{figure}[t!]
\begin{center}
\includegraphics[scale=0.6,angle=-90]{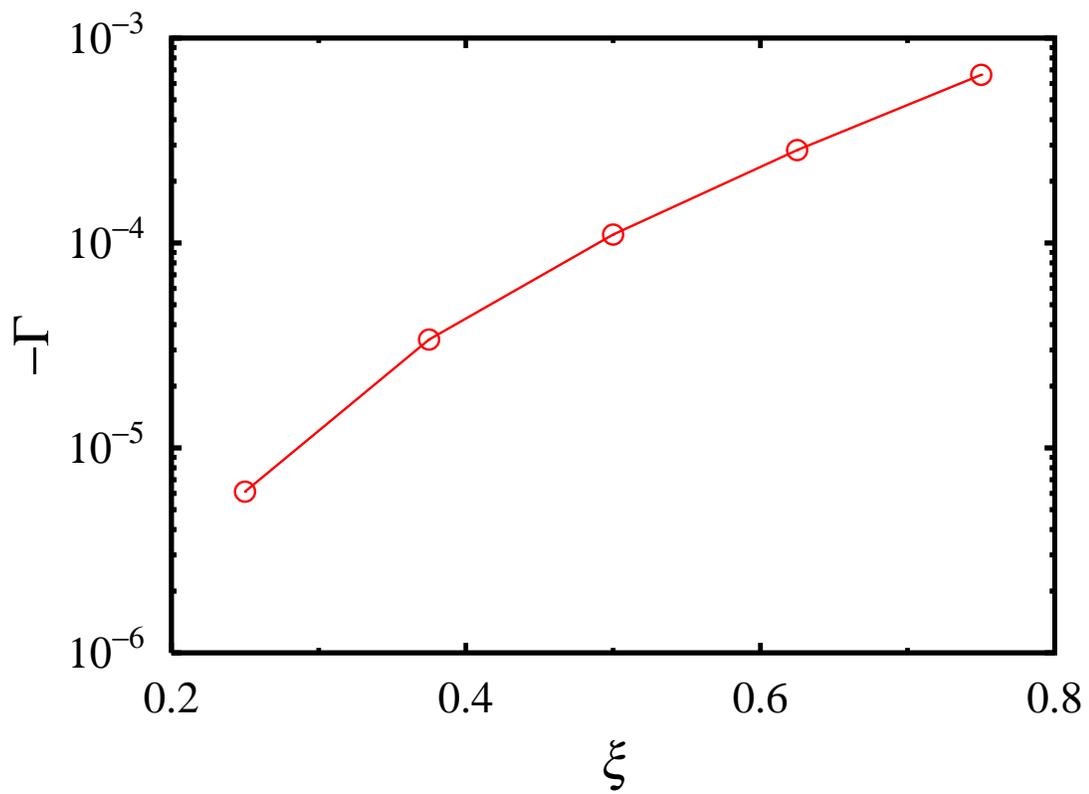}
\end{center}
\caption{The value of $\Gamma$ as defined by relation $\langle u \rangle/\langle \tilde{u} \rangle -1=\Gamma Re^4_{\tau}$ is shown for different values of $\xi$}
\label{fig:12}
\end{figure}

For the case of numerical results we concentrate on the case of transversal slip strips. To do so we will carry out numerical simulations with a $2d$ channel with dimensions $L_x$ (stream-wise) and $L_y$ (wall normal). The boundary condition of free shear is concentrated in a segment $H$ on both top and lower walls with a slip percentage $\xi$ immediately given by $\xi=H/L_x$ (see figure \ref{fig:01}).
The computations are made until the flow reaches a steady state. Monitoring the change of the kinetic energy budget and the time derivative terms is used to judge the steady state itself. \\
Figure \ref{fig:02} shows the time evolution of the averaged kinetic energy $\langle u^2_{x}+ u_y^2\rangle$ as a function of  time  for a friction Reynolds number $Re_{\tau}=2.245$, grid mesh $L_x=64$,$L_y=84$ and constant density $\rho=1$. LBM and FDM are used and in both cases with the same viscosity ($\nu=1/6$), driving pressure gradient ($-dP/dx=1.88 \times 10^{-6}$) and slip percentage $\xi=0.5$ . The same kind of plot is shown in the inset of figure \ref{fig:02} for a  higher friction Reynolds number $R_{\tau}=10.04$ and parameters $L_x=64$,$L_y=168$,$\nu=1/6$,$-dP/dx=4.72 \times 10^{-6}$,$\xi=0.5$. Let us notice that the agreement on the kinetic energy evolution is excellent with both schemes and it converges more rapidly towards a stationary state in the case of LBM due to the fact that in the kinetic scheme the Poisson equation for the pressure has not to be solved as in the case of FDM. In particular for the case with $Re_{\tau}=2.245$ the LBM is $5$ times faster and in the case with $Re_{\tau}=10.04$ is up to $15$ times faster.\\
In Figure \ref{fig:03} we show the velocity profiles along the channels for the case with $Re_{\tau}=2.245$ considered in figure \ref{fig:02}. In the inlet region the profile of Poiseuille type (parabolic profile with zero slip length) while in the middle of the slip region, as one can see, the local stress at the boundaries is zero and consequently the local viscous stress is minimized due to this free shear condition. As we can see not only global local properties (as it is the case of figure \ref{fig:02}) but also local properties from both numerical schemes agree very well.\\
In order to quantify the slip effects in the numerical simulations, we use the slip length $L_s$ as the distance from the wall at which the extrapolated velocity profile is zero. To do so, we use the volume averaged streamwise velocity
$$\langle u\rangle=\frac{1}{L_xL_y} \int u_x dx dy$$
and extrapolate at zero the boundary velocity.  In figure \ref{fig:05} the volume averaged slip length is shown as a function of the slip localization $\xi$ for a friction Reynolds number $Re=2.245$. The slip length has been obtained from steady state configurations with FDM and LBM and has been extrapolated using a first order accuracy and a fourth order accuracy.  From the numerical results, the agreement with the analytical estimate and the numerical results is quite satisfactory: small discrepancy between the two schemes is observed for $\xi$ approaching $1$ but the overall trend is very well reproduced by both.\\
Figure \ref{fig:06} shows the relative departure from the creeping flow solution $1-\langle u \rangle/\langle \tilde{u} \rangle$ for various friction Reynolds numbers. For small Reynolds number  $\langle u \rangle/\langle \tilde{u} \rangle$  is almost $1$ while, in increasing the Reynolds number it decreases due to the advective motion. Within the range of friction Reynolds numbers shown, no turbulent eddy is observed. If the whole wall were an homogeneous no slip,  $\langle u \rangle/\langle \tilde{u} \rangle$  would be independent of $Re_{\tau}$. On the other hand, when the mixed boundary condition (as it is this case) is switched on,  $\langle u \rangle/\langle \tilde{u} \rangle$ is dependent on $Re_{\tau}$, even in the laminar regimes under consideration.  According to our analytical investigation, for a given boundary condition (fixing $\xi$),  the drag modification is proportional to the fourth power of the friction Reynolds number and this can be verified explicitly from this plot. More precisely,  relation (\ref{master})  would imply
\be\label{master2}
1-\langle u \rangle/\langle \tilde{u} \rangle =  \langle ({\bm u}{\bm u}):({\bm \nabla}\tilde{\bm u})\rangle/\langle \tilde{u}\rangle  
\ee
and to verify the correctness of this relation, together with $1-\langle u \rangle/\langle \tilde{u} \rangle$   we plot also the right hand side of (\ref{master2}) and we observe an excellent agreement between the two contributions. \\
In figure \ref{fig:07} for the same set of parameters as figure \ref{fig:06} we show the relative change of the slip length with respect to its creeping flow solution. As expected, the same scaling behavior is observed and finite Reynolds effects become of the order of 10 percent soon after the friction Reynolds number is of the order $Re_{\tau} \sim 10$.\\
In figures \ref{fig:08} and \ref{fig:09}  the streamwise velocity is show for two Reynolds friction numbers $Re_{\tau}=2.245$ and $Re_{\tau}=10.04$ (the same as figure \ref{fig:01}) . While for the smallest $Re_{\tau}=2.245$ the profile is almost symmetric around the slip area, soon after $Re_{\tau}$ is increased an asymmetry, due to the advective flow is observed (see figure \ref{fig:09}). This asymmetry is responsible for the drag modification observed in the simulations and it develops from the creeping flow solution as a function of the Reynolds number.  For this case, in relation to a drag enhancement of ten percent the degree of asymmetry of the streamwise profile is of the same order.\\
In figure \ref{fig:11}, the relative change $1-\langle u \rangle/\langle \tilde{u} \rangle$ is considered for $\xi=0.25$, $\xi=0.5$ and $\xi=0.75$. The result is that using both LBM and FDM, excellent agreement is observed with respect to the functional behavior $\sim Re^4_{\tau}$ down to relative changes of $10^{-8}$. This is a stringent benchmark test for both numerical procedures for their degree of accuracy. \\
Finally, in figure \ref{fig:12}  the value of $\Gamma$, as extracted from the numerical simulations is shown as a function of $\xi$. As $\xi$ approaches $1$ the value of $\Gamma$ is higher due to the fact that the effect of the non linear terms is enhanced and non trivially triggered by the boundary condition.

\section{Conclusions}\label{sec6}

The importance of boundary slip velocity at finite Reynolds numbers  has been studied for open channel flows. The local and global properties of an incompressible Newtonian fluid with mixed boundary conditions of no-slip and free shear have been characterized as a function of the boundary geometry and Reynolds number. Although it would be easy to explain qualitatively the drag modification in the flow, its sensitivity with respect to the boundary condition and bulk properties may result a complicate subject. In our case, we have carried out numerical simulations with Lattice Boltzmann models (LBM) and Finite Difference methods (FDM) to highlight these properties and a  main role is played by non linear effects that are induced by the non homogeneity of the boundary condition. For small Reynolds numbers we have found interesting flow behaviors triggered by the boundary condition and very well controlled by the bulk Reynolds number. The overall physical picture is that a drag modification is induced by the non-uniform profile of the slip velocity at the boundary and it develops, for a given boundary realization, with a scaling law behavior as $Re^2$. The corresponding flow field is still laminar but looses its symmetry properties in the channel, with macroscopic effects up to the ten percent in the effective slip length. The interplay and the physical correctness  of both numerical schemes is then validated in the framework  of the theoretical predictions for slip conditions realized on strips and reveals a pronounced efficiency of LBM with respect to FDM. This is of particular interest especially when we move towards the study of the same kind of problems in $3d$ geometries with much more realistic boundary conditions of slip spots and slip patterns on heterogeneous planar surfaces.

\section*{Acknowledgments}

We acknowledge D.Lohse for constructive discussions and a critical reading of the manuscript.

\section{Appendix}

In this appendix we will detail the calculation leading to the analytical expressions used throughout the text. Let us assume  $0<Re_\tau\ll 1$ and expand ${\bm u}$ and $p$ with respect to $Re_\tau$ in the following form
\be
{\bm u}=
Re_\tau^1{\bm u}^{(0)}
+
Re_\tau^3{\bm u}^{(1)}
+
Re_\tau^5{\bm u}^{(2)}+...,
\ee
\be
p=
Re_\tau^0 p^{(0)}
+
Re_\tau^2 p^{(1)}
+
Re_\tau^4 p^{(2)}+...,
\ee
where the overscript $(0)$, 
indicate the creeping flow solution,  
\be
\tilde{\bm u}=Re_\tau{\bm u}^{(0)},
\ \ \ 
\tilde{p}=p^{(0)}.
\ee
All the even orders (i.e., $O(Re_\tau^{2n})$) of the velocity vector are identically zero because there is not any advection or driving forcing 
to balance with the divergence of the stress tensor 
containing the $O(Re_\tau^{2n})$ velocity. \\
The continuity equation is now expressed as
\be
{\bm \nabla}\cdot{\bm u}^{(n)}=0\ \ \ 
(n=0,1,...).
\ee
The $O(Re_\tau^0)$, $O(Re_\tau^2)$ and $O(Re_\tau^4)$
momentum equations (\ref{dimensionless2}) are respectively written as
$$
0=-{\bm \nabla} p^{(0)}+ \nabla^2{\bm u}^{(0)}+{\bm e}_x,
$$
$$
{\bm \nabla}\cdot({\bm u}^{(0)}{\bm u}^{(0)})
=-{\bm \nabla} p^{(1)}+\nabla^2{\bm u}^{(1)},
$$
$$
{\bm \nabla}\cdot({\bm u}^{(1)}{\bm u}^{(0)}
+{\bm u}^{(0)}{\bm u}^{(1)})
=-{\bm \nabla} p^{(2)}+\nabla^2{\bm u}^{(2)}.
$$
For $n=0, 1$ and $2$,  the boundary conditions are given by
$$
{\bm u}^{(n)}=0\ \ \ {\rm on}\ \ \ {\rm NSW},
$$
$$
({\bm n}\cdot{\bm S}^{(n)})\times {\bm n}=0
\ \ \ {\rm on}\ \ \ {\rm FSW},
$$
where 
${\bm S}^{(n)}=\frac{1}{2}\left\{
{\bm \nabla}{\bm u}^{(n)}+({\bm \nabla}{\bm u}^{(n)})^{\rm T}
\right\}$. With these expansions, starting from equation (\ref{master}),
we obtain the following relation
\begin{equation}\label{P1}
\begin{split}
\langle u\rangle-\langle \tilde{u}\rangle
=&
Re_\tau^3\langle u^{(1)}\rangle
+
Re_\tau^5\langle u^{(2)}\rangle
\\=&
Re_\tau^3
\langle ({\bm u}^{(0)}{\bm u}^{(0)}):
({\bm \nabla}{\bm u}^{(0)})\rangle
+
Re_\tau^5
\langle ({\bm u}^{(1)}{\bm u}^{(0)}):
({\bm \nabla}{\bm u}^{(0)})\rangle
+
Re_\tau^5
\langle ({\bm u}^{(0)}{\bm u}^{(1)}):
({\bm \nabla}{\bm u}^{(0)})\rangle.
\end{split}
\end{equation}
The first and second terms in the rhs of equation (\ref{P1})
are zero because  the kernel of the volume integral is written in the divergence form  which can be rewritten using the area integral with 
no wall-normal flux:
\begin{equation}
\begin{split}
\int_{\mbox{\footnotesize fluid}}\!\!\!{\rm d}^3{\bm x}\ 
({\bm u}^{(0)}{\bm u}^{(0)}):
({\bm \nabla}{\bm u}^{(0)})
=&
\int_{\mbox{\footnotesize fluid}}\!\!\!{\rm d}^3{\bm x}\ 
{\bm \nabla}\cdot\left\{
{\bm u}^{(0)}({\bm u}^{(0)}\cdot{\bm u}^{(0)})/2
\right\}
\\=&
\oint_{\mbox{\footnotesize wall}}\!\!\!{\rm d}^2{\bm x}\ 
\underbrace{
({\bm n}\cdot{\bm u}^{(0)})
}_{=0}
({\bm u}^{(0)}\cdot{\bm u}^{(0)})/2,
\end{split}
\end{equation}
\begin{equation}
\begin{split}
\int_{\mbox{\footnotesize fluid}}\!\!\!{\rm d}^3{\bm x}\ 
({\bm u}^{(1)}{\bm u}^{(0)}):
({\bm \nabla}{\bm u}^{(0)})
=&
\int_{\mbox{\footnotesize fluid}}\!\!\!{\rm d}^3{\bm x}\ 
{\bm \nabla}\cdot\left\{
{\bm u}^{(1)}({\bm u}^{(0)}\cdot{\bm u}^{(0)})/2
\right\}
\\=&
\oint_{\mbox{\footnotesize wall}}\!\!\!{\rm d}^2{\bm x}\ 
\underbrace{
({\bm n}\cdot{\bm u}^{(1)})
}_{=0}
({\bm u}^{(0)}\cdot{\bm u}^{(0)})/2.
\end{split}
\end{equation}
The third term in the rhs of (\ref{P1}) is the only one that can take can take a non-zero value.  Therefore, we obtain
$$
\langle u\rangle-\langle \tilde{u}\rangle
\rightarrow 
Re_\tau^5
\left
\langle 
{\bm u}^{(1)}\cdot
\left\{
({\bm u}^{(0)}\cdot{\bm \nabla}){\bm u}^{(0)}y
\right\}
\right\rangle
\ \ \ 
{\rm as}\ \ \  Re_\tau\rightarrow +0.
$$
It should be noted that for the expression 
$$\langle u\rangle-\langle \tilde{u}\rangle =\Gamma Re_\tau^5,$$ 
the prefactor 
$$\Gamma =
\left
\langle 
{\bm u}^{(1)}\cdot
\left\{
({\bm u}^{(0)}\cdot{\bm \nabla}){\bm u}^{(0)}
\right\}
\right\rangle$$ 
is dependent upon the perturbed velocity component
but independent of the Reynolds number.

\newpage

\end{document}